\documentclass{pasj00}

\usepackage{graphicx}
\usepackage{tabularx}
\usepackage{here}

\draft

\begin{document}
\SetRunningHead{Author(s) in page-head}{Running Head}

\title{\bf Time-evolution of Peak Energy and Luminosity Relation within Pulses 
for GRB 061007: Probing Fireball Dynamics}

\author{Masanori \textsc{Ohno}$^1$, Kunihito \textsc{
Ioka}$^2$, Kazutaka \textsc{Yamaoka}$^3$, \\
Makoto \textsc{Tashiro}$^4$, Yasushi \textsc{Fukazawa}$^5$, and Yujin
E. \textsc{Nakagawa}$^6$}%

\affil{$^1$ Institute of Space and Astronautical Science, \\Japan
Aerospace Exploration Agency (ISAS/JAXA), \\3-1-1 Yoshinodai, Sagamihara,
Kanagawa 229-8510, Japan}

\email{ohno@astro.isas.jaxa.jp}

\affil{$^2$ Theory Division, KEK (High Energy Accelerator Research
Organization), \\1-1 Oho, Tsukuba 305-0801, Japan}

\affil{$^3$ Department of Physics and Mathematics,  Aoyama Gakuin
University, \\5-10-1, Fuchinobe, Sagamihara 229-8558}

\affil{$^4$ Department of Physics, Saitama University, \\255 Shimo-Ohkubo, Sakura, Saitama 338-8570}

\affil{$^5$ Department of Physical Sciences, School of Science, Hiroshima
 University \\ 1-3-1 Kagamiyama, Higashi-Hiroshima, Hiroshima 739-8526}

\and

\affil{$^6$ Institute of Physical and Chemical Research (RIKEN),
 2-1 Hirosawa, Wako, Saitama 351-0198}


\maketitle

\begin{abstract}

We perform a time-resolved spectral analysis of bright, long Gamma-ray
 burst GRB 061007
 using Suzaku/WAM and Swift/BAT. Thanks to the large effective area of
 the WAM, we can investigate the time evolution of the spectral peak
 energy, $E^{\rm t}_{ \rm
 peak}$ and the
luminosity $L^{\rm t}_{\rm iso}$ with 1-sec time resolution, and we find that
 the time-resolved pulses also satisfy the $E_{\rm
 peak}-L_{\rm iso}$ relation, which was found for the time-averaged
 spectra of other bursts, suggesting the same physical conditions in
 each pulse. Furthermore, the initial rising
 phase of each pulse could be an outlier of this relation with higher
 $E^{\rm t}_{\rm peak}$ value by about factor 2. This difference could suggest that
the fireball radius expands by a factor of $2-4$ and/or bulk Lorentz
 factor of the fireball is decelerated by a factor of $\sim$ 4 during
 the initial phase, providing a new probe of the fireball dynamics in
 real time.
 
\end{abstract}

\section{Introduction}

Many characteristics of Gamma-ray Bursts (GRBs) have been
revealed by the previous observations, and now it is widely believed that
GRBs are the one of the most powerful explosion in the Universe, 
originating at cosmological distances. The association with the
energetic supernovae was found for some long duration GRBs. This 
provides a strong evidence that the progenitor of the long GRBs is the
core collapse of a massive star. The observed prompt gamma-ray
spectra are often fitted well by a smoothly connected broken 
power-law called as the Band function (\cite{Band 1993}).
However, we are still not able to uniquely associate the measured properties to 
 the radiation mechanism of the prompt gamma-ray emission. 

Although, the observed behavior of GRBs, such as the
intensity and the peak energy of the $\nu F _\nu$ spectrum ($E_{\rm
peak}$) are
dramatically different from burst to burst, some clear correlations
between these parameters were reported, such as the correlation between 
isotropic equivalent radiation energy $E_{\rm iso}$ and the peak energy
in  the rest frame of the burst ($E_{\rm
peak,src}$) (\cite{Amati 2002}; \cite{Amati 2006}), or the isotropic
equivalent peak luminosity $L_{peak,\rm iso}$ and the $E_{\rm
peak,src}$ (\cite{Yonetoku 2004}), and the total radiation energy corrected
for the jet opening angle and the $E_{\rm
peak,src}$ (\cite{Ghirlanda 2004b}). These relations are often used to
estimate the inferred redshift value, and Yonetoku et al. (2003) estimated
redshift for large BATSE sample and found that some of them have large
redshift with z $>$ 10. Such high-z GRBs are important to constrain 
the cosmological parameters.
Furthermore, the same kind of correlation have been reported not
only for each burst but also for individual pulses of a burst 
internally. Several authors studied this hardness-intensity 
correlation (HIC) for time-resolved spectral analysis. Golenetski et
al. 1983 found a power-law correlation between observed energy flux and
the spectral characteristic energy E$_0$, and the power-law correlation 
index was
found to be a typical value of 1.5-1.7. Kargatis et al. (1994) confirmed
this correlation but the power-law index showed a larger spread. They obtained the
power-law index of 2.2$\pm$1.0. 
Strohmayer et al. (1998) also found that this
power-law correlation is valid using Ginga data. Borgonovo \& Ryde (2001)
also studied this correlation by strong BATSE sample. They found the 
specific feature that there is the sharp transition from one power-law 
track to the second track with the
same power-law index in the decay phase of the pulse. They explained
this track-jump behavior as the transition to a new pulse. 
The fact that the HIC holds for both each pulse and within the
pulse implies that the same physical condition is realized in
each pulse. Therefore, to investigate this relation it 
is very important to consider the radiation mechanism of the GRB prompt
emission. Recently, Liang et al. (2004) 
analyzed the $L_{\rm iso}$ and $E_{\rm peak,src}$ of 
the 2408 time-resolved spectra for 91 BATSE GRBs, assuming that the 
burst rate as a function of redshift is proportional to the star
formation rate. Yoshida et al. (2007) measured the $L_{\rm iso}$ and 
$E_{\rm peak,src}$ value for time-resolved spectra of HETE-2
bursts. Both of them found that 
$L_{\rm iso} \propto E_{\rm \rm peak,src}^2$, and Yoshida et al. (2007)
found that this relation is tighter for each spike-averaged spectra
rather than that of simply 5.2 s-divided spectra. 
Therefore, this relation could be different for finer time-resolved
spectra. 
However, most of previous
analysis was done using data for each pulse or for the decay phase of such 
pulses, while the detailed properties of time-resolved spectra, including 
initial rising phase for each pulse have not been considered and thus 
are still ambiguous.
In this paper, we report on the time-resolved spectroscopy of the bright
long GRB 061007 observed by the Suzaku/WAM and Swift/BAT.  
 This burst was also observed by many ground-based telescopes and
the redshift was measured to be 1.261  (\cite{Osip 2006};
\cite{Jakobsson 2006}), and Schady et al. (2007) suggested that this burst
has an early jet break time which is within 80 s of the prompt emission,
and a highly collimated outflow is likely case for this
burst. Using this bright burst with a measured redshift, we can 
investigate the relation between $L_{\rm iso}$ and $E_{\rm
peak,src}$ within each pulse, including both the pulse-rise and the decay
phase.  This is possible since 
 Suzaku/WAM has the largest effective area from
300 keV to 5 MeV than any previous missions. Furthermore, since Swift/BAT
covers an energy range from 15 keV to 150 keV, the joint analysis using 
Suzaku/WAM and Swift/BAT gives us further constraints on the spectral
parameters. Therefore, this joint, broad-band Suzaku/WAM and Swift/BAT analysis
allows the excellent time-resolved spectral analysis with both fine
time resolution and high sensitivity.  We also investigate the
time-averaged properties of this burst such as $E_{\rm iso}$ -- $E_{\rm
peak,src}$  (Amati) relation and $L_{\rm iso}$ -- $E_{\rm peak,src}$
(Yonetoku) relation in order to compare with other bursts. Thus, we
denote our ``time-resolved''  peak energy and luminosity 
 relation as  $E^{\rm t}_{ \rm peak,src}$ -- $L^{\rm t}_{ \rm iso}$  relation in this
paper to distinguish from those time-averaged relationships.

\section{Observations}

\subsection{Suzaku/WAM}

The Suzaku/WAM (WAM) is the active shield of the Hard X-ray detector
(HXD-II) (\cite{Takahashi 2006}, \cite{Kokubun 2006})
onboard Suzaku (\cite{Mitsuda 2006}). 
It consists of large and thick BGO crystals  
and it is also designed to monitor the all-sky flux over the 50 keV to 5 MeV 
band. It has the largest effective area from 300 keV to 5
MeV of all GRB missions and thus enables us to perform a
wide-band spectroscopy of GRBs with a high sensitivity (\cite{Yamaoka
2005}, \cite{Yamaoka 2006}). There is still uncertainty in the detector 
response matrices of the WAM and the observed flux fluctuates by 10-20\% on average and about 40\% at the maximum depending on
the incident azimuthal ($\phi$) and zenith ($\theta$) angle of GRBs
(\cite{Ohno 2005}). We found that this uncertainty (as
compared to the pre-flight calibration) is mainly caused 
by the absorption by the structures attached inside the satellite
panel. We confirmed that this trend is
the same even in the in-orbit environment by the cross-calibration between
Swift/BAT and Konus-Wind, using the data of simultaneously detected
GRBs (Sakamoto et al. 2008 in preparation).

The WAM outputs two data types, the transient (TRN) data
and the gamma-ray burst (GRB) data. The TRN data are always accumulated
with 1 sec time resolution and 55 energy channels. This can be used to
monitor the bright soft gamma-ray sources with the Earth
occultation method, as was done by the CGRO/BATSE (\cite{Ling 2000}).
On the other hand, the GRB data are recorded for 64 sec only when
the GRB trigger is activated, and the data cover 8 sec before and 56
sec after the trigger time. The GRB data have four
energy channels with 1/64 sec time resolution, in addition
to the spectral data in 55 pulse height channels with 0.5 sec time
resolution. 

\begin{figure}[t]
\begin{center}
\rotatebox{-90}{\resizebox{6cm}{!}{\includegraphics{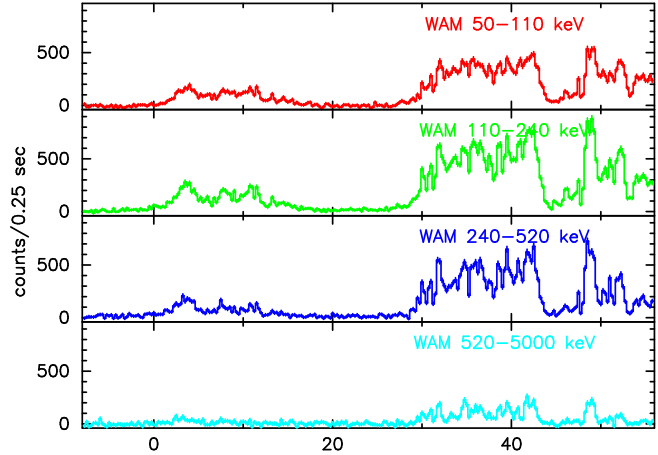}}}
\caption{Light curve of GRB 061007 obtained by the WAM GRB data in 1/64
 s time resolution. This
 light curve is divided into each energy band of the GRB data (TH0:
 50-110 keV, TH1:110-240 keV, TH2:240-520 keV, and TH3:520-5000 keV).}
\label{grb061007_wamgrblc}
\end{center}
\end{figure}

GRB061007 triggered the WAM at 10:08:05 UT, October 07, 2006
(T$_0$(WAM)). Three
large peaks as well as a following smaller peak were clearly seen in the
WAM light curve as shown in figure \ref{grb061007_wamgrblc}. Gamma-ray photons are strongly
detected by the WAM-3 detector. The T$_{\rm 90}$ duration of this burst
was about 59 seconds in the 50-5000 keV band (\cite{Yamaoka 2006b}). 
Since this burst was so
long that the emission lasted more than 56 s after the WAM trigger, the
GRB data system could not record the whole emission profile. This is why the
light curve of the WAM GRB data look truncated above 56 s (see Figure 2). Based
on the position information of 
this burst provided by the Swift as mentioned below,
the incident direction against the WAM detector was found to be
($\theta$, $\phi$) = (92\degree, 198\degree). This incident direction is consistent
with the strong detection by the WAM-3 detector, and the detector
response of this direction is known to be reliable by the pre-flight
and in-orbit calibrations. The uncertainties of the response matrix
should be within 20-30\% for this burst.

\subsection{Swift/BAT}

The Swift/BAT (BAT) was also triggered by GRB 061007, 3 s after the
WAM trigger, at 10:08:08 UT, October 07,
2006. The refined BAT position for this burst was RA = $\timeform{03h05m11s.8}$
 Dec = $\timeform{-50D29'47.7''}$ (J2000) with an uncertainty of 0.9$\arcmin$ at 90\% containment (\cite{Markwardt 2006}). The BAT light curve showed a
multi-peaked structure with a duration of about 100 seconds. The
Swift/XRT and Swift/UVOT observed this burst 80 s and 400 s after the
BAT trigger, respectively, and they found a bright fading source.

%

\section{Data Analysis and Results}

Although, the main purpose of this study is to investigate the
behavior of the time-resolved spectra of GRB 061007, before we 
perform the time-resolved spectral analysis, the time-averaged spectral 
parameters have been also investigated. 
We utilize the transient (TRN) data for the WAM because GRB data of
the WAM do not contain the whole emission of this burst due to its
long duration. Therefore, we can analyze only 1-sec time resolution data
for this analysis. We perform the spectral analysis by applying three spectral models.
The first one is a simple
power law (PL) model:

\begin{equation}
N(E)=A \times \Biggl(\frac{E}{100 \rm keV}\Biggr)^{\alpha}
\end{equation}

\noindent where  A is the
normalization constant at 100 keV in photons cm$^{-2}$ s$^{-1}$ keV $^{-1}$, and $\alpha$ is the power law photon
index.

The second model is a power law with an exponential cutoff (CPL):

\begin{equation}
N(E)=A \times \Biggl(\frac{E}{100 \rm keV}\Biggr)^{\alpha} {\rm exp}\Biggl(-\frac{E(2+\alpha)}{E_{\rm peak}}\Biggr)
\end{equation}

\noindent where $E_{\rm peak}$ is the peak energy in the $\nu F_\nu$ spectrum and
it represents the energy at which most of the power is emitted.

The third model is a smoothly connected broken power law model known as
the Band model (\cite{Band 1993}).

\begin{eqnarray}
N(E) & = & A \times \Biggl(\frac{E}{100 \rm keV}\Biggr)^{\alpha} {\rm exp}
 \Biggl(-\frac{E(2+\alpha)}{E_{\rm peak}}\Biggr),\nonumber\\
& & {\rm for}~ E <
 \frac{(\alpha-\beta)E_{\rm peak}}{(2+\alpha)}\nonumber\\
     &   & A \times \Biggl(\frac{E}{100 \rm keV}\Biggr)^{\beta} \Biggl[\frac{(\alpha - \beta) E_{\rm
      peak}}{100 \rm keV(2+\alpha)}\Biggr]^{(\alpha-\beta)} {\rm exp}(\beta -
      \alpha), \nonumber\\
& &{\rm for}~ E \geq
 \frac{(\alpha-\beta)E_{\rm peak}}{(2+\alpha)}
   \end{eqnarray}

\noindent where $\alpha$ is the power law photon index in the lower energy band,
and $\beta$ is that in higher energy band.
In addition, in these basic models, we add the constant factor when we
perform the joint spectral analysis between the WAM and the BAT to take
into account for the uncertainties of the detector response matrix of
the WAM. Throughout this analysis, we fit the spectrum from 120 keV to 5000
keV for the WAM and from 13 keV to 150 keV for the BAT, respectively.

In order to extract the dead-time corrected light curve and the spectrum,
we use the standard FTOOLS; {\tt hxdmkwamspec} and {\tt
hxdmkwamlc} for transient data, and {\tt hxdmkbstlc} and {\tt
hxdmkbstspec} for GRB data, which are included in the
HEAsoft software package version 6.0.6
(http://heasarc.gsfc.nasa.gov/lheasoft/). We also apply a 2\% systematic
error to the WAM spectrum only for the time-averaged spectral analysis
due to its high counting rate. 
 We also derive the light curve and the spectra from Swift/BAT data for joint
spectral analysis. When we extract the spectrum or light curve
from the BAT event data, we use the standard FTOOLS; {\tt
batbinevt}, after we re-processed the energy
calibration, making the detector quality map and  mask weighting, using
the latest version of the software and calibration data base.
We also apply systematic error for the BAT spectra, using 
{\tt batphasyserr} script.
The systematic error vector is retrieved from the calibration data base.

We use the XSPEC version 11.3.2 for spectral analysis
(\cite{Arnaud 1996}). All quoted error are 90\% confidence level.

\subsection{Time-averaged Analysis}

\begin{figure}[t]
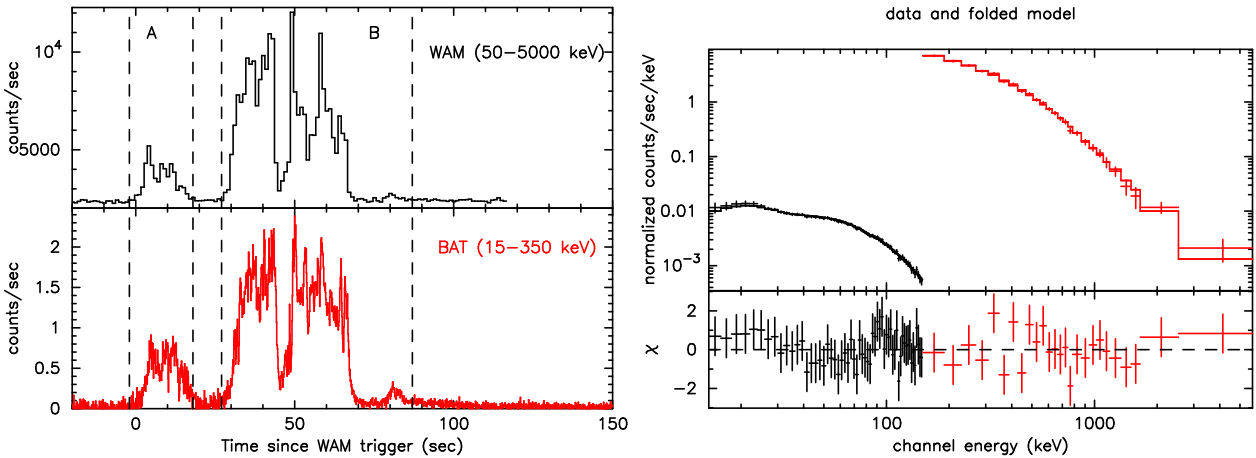

\begin{center}
\hspace{-0.8cm}
\rotatebox{-90}{\resizebox{6cm}{!}{\includegraphics{figure/figure2a.ps}}}
\rotatebox{-90}{\resizebox{6cm}{!}{\includegraphics{figure/figure2b.ps}}}
\caption{Top: Light curve of GRB 061007 obtained by the WAM monitor data (TRN
 data; top) and the BAT (bottom). 
The time regions for the time averaged spectral analysis are also
 shown by the vertical dashed lines. Bottom: Joint spectral fitting with a Band model for the WAM (red) and
 the BAT (black) data for the data A plus B. 
The solid line represent the best fit Band model.}
\label{grb061007_trnlcandspec}
\end{center}
\end{figure}

We can see that the emission profile is well separated into first weak FRED
 like structure and second bright, multi-peaked episode.
 Therefore, firstly, we divide this burst into two time regions. 
The first time region named
as interval A is selected from T$_0$(WAM)-2.0 s to T$_0$(WAM)+18.0 s,
 where T$_0$(WAM) means the WAM
triggered time, and the second time
region of the interval B is from T$_0$(WAM)+27.0 sec to
T$_0$(WAM)+87.0 sec. We also analyze the integrated time region from
T$_0$(WAM)-2.0 sec to T$_0$(WAM)+85.0 sec.  We also extract the BAT spectra from the same time
interval as that of the WAM in order to perform the joint spectral
analysis. 
The selected time region is shown in the light curve of the WAM and BAT 
in figure
\ref{grb061007_trnlcandspec}. The best-fit
parameters obtained by the joint analysis for each time region are shown
in Table \ref{bestfit_timeavegrb061007}.  Figure \ref{grb061007_trnlcandspec}
shows the spectra of the joint analysis with the
WAM and the BAT for whole time region (A+B). 
We find that the spectrum of this burst can be described
by the typical Band model for each time interval. We
 obtain the photon index of low-energy part, $\alpha$ of
 $-0.91 (\pm 0.07), -0.73 (\pm 0.03),
 -0.79 (\pm 0.03)$, photon index of
 high-energy part, $\beta$ of $-3.00^{+0.41}_{-1.45},
 -3.22^{+0.18}_{-0.26}, -3.21^{+0.20}_{-0.30}$, 
 and the peak energy, $E_{\rm peak}$ of 354$^{+28}_{-26}$ keV,
 468 ($\pm 10 $) keV, 460 ($\pm 10$) keV, for interval A, B, and A+B, 
respectively. These spectral parameters are consistent with that of
 reported by Konus-Wind (\cite{Golenetskii 2006}). The 100-1000 keV fluence and the peak flux in
 1-s time scale is measured to be 1.6 ($\pm 0.15$) $\times 10^{-5}$ erg
 cm$^{-2}$ and 14.5 ($\pm 0.9$) photons s$^{-1}$
 cm$^{-2}$, respectively.
The constant factor of the
WAM against the BAT is about 1.10 for any time regions. 
These values are consistent with the
 current uncertainties of the response matrix of the WAM. 
 Furthermore, we find that the spectral parameters can be tightly 
constrained by the joint analysis between the WAM and the BAT.

We also investigate other spectral properties of this burst using the
spectrum of the whole time interval (A+B).  First, we measure the
hardness ratio between 100-300 keV and 50-150 keV, using the
fluence ratio based on the spectral fitting result and we obtain the
value of 3.52 ($\pm 0.13$). We plot this value on the
hardness-duration plane in Figure \ref{grb061007_HRandAmati} 
(left panel). We also plot the BATSE results
together for comparison, and we find that this burst locate at the standard
long/soft regime like BATSE results. Then, we investigate the total
emitting energy of this burst.  We calculate the isotropic equivalent
total energy $E_{\rm iso}$ from 1.0 to 10000 keV and the peak energy at 
the rest frame $E_{\rm peak, src}$, using cosmological parameters 
(H$_0$, $\Omega_{\Lambda}$, $\Omega_m$) = (65, 0.7, 0.3), and we obtain
the $E_{\rm iso}$ of 1.0$(\pm 0.3) \times 10^{52}$ erg, and the $E_{\rm
peak, src}$ of 1040$^{+23}_{-22}$ keV, respectively. We plot 
this result in the $E_{\rm peak, src}$-$E_{\rm iso}$ plane and we show 
it in Figure \ref{grb061007_HRandAmati} right.  
For long GRBs, it is well known that there is a strong
correlation between  $E_{\rm iso}$ and  $E_{\rm
peak, src}$  (Amati relation; \cite{Amati 2002}, \cite{Amati 2006}). Thus,
we also plot the previous results reported by \cite{Amati 2006} in the 
same figure, and we find that 
the $E_{\rm iso}$ and $E_{\rm peak, src}$ value of GRB 061007 is quite
consistent with Amati relation. Figure \ref{grb061007_yonetoku} also
shows a comparison of the
$E_{\rm iso}$ and 1-sec peak luminosity (Yonetoku) relation
(\cite{Yonetoku 2004}), and we find that GRB 061007 also satisfies the
Yonetoku relation.
From those time-averaged spectral analysis, 
we conclude that there is no large systematic
problem in our analysis and that GRB 061007 is a typical long GRB with
many common characteristics such as the duration, the spectral hardness, 
and the total emitting energy.

\subsection{Time-resolved Analysis}

After we confirmed that there was no problem in the time-averaged
spectral analysis between the WAM and the BAT, we investigate the 
time-resolved spectral properties of this burst. In this analysis, we divide
the WAM data in 1.0 s time resolution. We also divided the BAT data
for the same time region. We extract total 58 time-resolved
spectra and perform the spectral analysis with the same manner as that
we used in the time-averaged analysis. The time-resolved spectra are
also described by the CPL or Band model. We can constrain the high
energy photon index $\beta$ from only a few spectra which belong to 
the bright time region.  Therefore, we applied the fixed value of
$\beta$ of $-3.2$ which is obtained at the time-averaged spectral
analysis. 
The best-fit parameters for 58-time-resolved 
spectra are shown in Table \ref{grb061007_timeresolvedpar} and 
\ref{grb061007_timeresolvedpar2}.
From these time-resolved analysis, we find the evolution of the
spectral shape during the burst. Figure \ref{grb061007_ufu}
shows examples of the $\nu F_\nu$ spectra with the best-fit 
Band model for various intensity region, and the change of the spectral
shape around the $E_{\rm peak}$ is clearly seen. 
In figure \ref{grb061007_Bandlcpar}, we present the
best-fit parameters obtained by the Band model fit as a function of time,
and both the low energy photon index $\alpha$ and the $E_{\rm peak}$ show a
hard-to-soft tracking behavior correlated with the burst intensity. 
The low energy photon index $\alpha$ changes between  $-1.5$ to $-0.5$ 
with an average value of $-0.84$, which is consistent with the time 
integrated parameter of
$-0.79 (\pm 0.03)$. 
The $E_{\rm peak}$ also evolves within the burst, and it
moves from 120 keV to 633 keV with an average value of 375 keV. 
Figure \ref{Epalpha} shows the relation between low energy photon index,
$\alpha$ and the $E_{\rm peak}$, and we find that only one time-resolved
spectrum have the low energy photon index of flatter than $-2/3$, which
is not allowed by the standard synchrotron shock scenario.

\section{Discussion}

\subsection{$E^{\rm t}_{ \rm peak}$--$L^{\rm t}_{ \rm iso}$ relation for time-resolved spectra}

 From time-resolved spectral analysis, we find that the time-resolved
  $E_{\rm peak}$; $E^{\rm t}_{ \rm peak}$
  changes are correlated with the burst intensity. In order to confirm this,
  we perform a correlation analysis between $E^{\rm t}_{ \rm peak}$ and time
  resolved luminosity; $L_{\rm
  iso}$ for time-resolved spectra. The $L^{\rm t}_{ \rm iso}$ value is measured 
  as the 1-sec time-averaged value for each of the 
  time-resolved spectra in 30 keV to 10000 keV energy range. 
  Figure \ref{EpLiso} 
shows the relation between
  $E^{\rm t}_{ \rm peak}$ and $L^{\rm t}_{ \rm iso}$. We can see a clear positive
  correlation between these two parameters. To quantify this, we
  calculate the Spearman rank-order coefficients ($r_s$), and we obtain
  the correlation coefficient of 0.78. This
  coefficient corresponds to the significance probability that there
  is no correlation of 5.4 $\times$ 10$^{-13}$. 
 Therefore, we confirm that the correlation surely exists at $> 3
  \sigma$ confidence level. We then fit this correlation by the simple
  power-law model like the Hardness-Intensity correlation, 
  the Amati relation, and the Yonetoku relation. The fit gives a best
  fit parameter as, 
\begin{equation}
 \Biggl(\frac{E^{\rm t}_{ \rm peak}}{1 ~\rm keV} \Biggr) = 492(\pm 24) \times
 \Biggl(\frac{L^{\rm t}_{ \rm iso}}{10^{52} ~\rm {erg~ s^{-1}}}
 \Biggr)^{0.43(\pm 0.03)}.
\label{epliso_alldata}
\end{equation}

The power-law index slightly flatter than the value obtained by
Golenetskii et al. 1983 but roughly agrees with the previous
analysis by Yonetoku 2004; $E_{\rm peak} \propto L_{\rm
iso}^{0.5}$.
Furthermore, our data set can constrain both the power-law index and the
normalization factor as  well as the previous analysis
 using only one burst data set. The $\chi^2$ value for this fit is 83/56
 and we still see a certain deviation in this relation around $L^{\rm t}_{ \rm
iso}/10^{52} ~{\rm erg ~s^{-1}} = 2 - 5$ (see Figure 8), and
the data point which have large dispersion from this relation tend to 
have a higher $E^{\rm t}_{ \rm peak}$ value by a factor of 2 compared with 
many other data. We calculate the $E^{\rm t}_{ \rm peak}/ (L^{\rm t}_{ \rm iso})^{0.43}$ for each time
region in order to separete these data point from other data which
follow the equation(\ref{epliso_alldata}). 
If the correlation follows  
$E^{\rm t}_{ \rm peak} \propto L^{\rm t~0.43}_{\rm iso}$ 
for any time region, this parameter should 
retain constant. Figure \ref{EpoverLiso045} shows the
$E^{\rm t}_{ \rm peak}/ (L^{\rm t}_{ \rm iso})^{0.43}$  parameter as a function of
time. We also show the best-fit constant value with
3 $\sigma$ confidence region. The 8 data points (T=4,5,6,7,31,32,48,50) 
 exceed 3 $\sigma$
limit of the constant value, and thus we consider these data points as
the outliers on the  $E^{\rm t}_{ \rm peak}-L^{\rm t}_{ \rm iso}$ plane. 
 We confirm that these outliers with large dispersion from the
main population are not caused by the systematic
uncertainties in the analysis procedures such as the uncertainties 
of detector response matrix, systematic effect to the obtained spectral
parameters, and the effect of the intense spectral evolution during the
initial rising phase. The details of this verification processes are
shown in the Appendix.

We then divide the derived parameter pairs into the main population,
where the $E^{\rm t}_{ \rm peak}/ (L^{\rm t}_{ \rm iso})^{0.43}$ correlation is acceptable, and the
outlier population, where 
that relation is broken at 3 $\sigma$ significance. We calculate
the correlation coefficient and perform the power-law fit for each
population separately, and obtain the coefficients of 0.90 ($P_{rs}$ =
6.9 $\times 10^{-18}$) for the main population, and 0.62 for the
outliers. 
The correlation coefficient of the
main population is marginally improved when we exclude the outliers. 
The outliers also have a certain degree of the correlation.
The power-law best-fit parameters for those two populations are found as follows,

\begin{eqnarray}
\Biggl(\frac{E^{\rm t}_{ \rm peak}}{1 ~\rm keV}\Biggr) &=& 456(\pm 25) \times
 \Biggl(\frac{L^{\rm t}_{ \rm iso}}{10^{52} ~\rm {erg~ s^{-1}}}
 \Biggr)^{0.46(\pm 0.03)} ~~~~~~~~~~ {\rm for~~ main~~ population}
 \nonumber \\
                                      &=&  909(\pm 147) \times
 \Biggl(\frac{L^{\rm t}_{ \rm iso}}{10^{52} ~\rm {erg~ s^{-1}}}
 \Biggr)^{0.25(\pm 0.14)}  ~~~~~~~~~~ {\rm for~~ outliers.}
\label{EpLisorelation}
\end{eqnarray}

For the main population, the best-fit parameter of the power-law fit does
not change significantly from that we derived from the entire data. 
We find that the power-law model is still valid for this relation.
Furthermore, this sparation also led to a very interesting result. 
From figure \ref{EpoverLiso045}, we can see that
all of these outliers correspond to the
rising phase of both initial weak peak and the second and third intense
pulses. This indicates that if we exclude the initial rising phase of
each pulse, we obtain much tighter $E^{\rm t}_{ \rm
peak,src}$--$L^{\rm t}_{ \rm iso}$ correlation for all GRB.
Therefore, it is probably more appropriate to 
exclude initial rising phase of each pulse when we 
use this relation as the redshift indicator like time-averaged $E_{\rm
peak,src}$ -- $L_{\rm iso}$ (Yonetoku) relation.
Moreover, it is likely that these
outliers, in other words initial rising phase of each pulse have a 
different correlation with higher normalization and
possibly smaller power-law index. However, we cannot give a strong
conclusion for this
because this power-law index varies, strongly depending on the criteria 
between the main population and outliers. Furthermore, we cannot
constrain the power-law index of outliers due to the 
small number of data. Therefore, the robust conclusion is that the
initial rising phase of burst have a different
$E^{\rm t}_{ \rm peak}-L^{\rm t}_{ \rm iso}$ correlation with higher $E^{\rm t}_{ \rm peak}$ value
from that of decay phase of burst. Note that this argumentation do not
form a circular logic because the main point of this study is not to 
improve the fit of  $E^{\rm t}_{ \rm peak}-L^{\rm t}_{ \rm iso}$
correlation by removing the outliers but to just find that these
outliers correspond to the initial phase of each pulse.   
This kind of trend that the initial phase of long GRBs has harder
spectrum than that of decay phase is already reported by \cite{Ghirlanda
2004}. However, the WAM data revealed that this harder-rising phase of long
GRBs also follow the another power-law relation with that of softer-decay
phase.   This difference between these different phases affects mainly the 
normalization of the relation. This fact can be useful to constrain 
the emission mechanism or dynamics of the emission site of the prompt 
emission of GRBs. 
 
\subsection{Implication for the emission radius and the bulk Lorentz factor}

Let us consider theoretical implications of our results
that the spectral peak energy $E_{\rm peak}$ evolves from hard to soft
in the rising phase of each pulse.
Although the emission mechanism of prompt GRBs is still enigmatic,
the leading models are (1) the synchrotron shock model
\citep{zm02}
and (2) the photosphere model \citep{mesree00, th07,ioka07}.

In the synchrotron shock model,
the peak energy is identified with the typical synchrotron energy
$E_{\rm peak}=\Gamma \hbar \gamma_m^2 e B/m_e c$ of electrons
that are shocked in the relativistic outflow with a Lorentz factor $\Gamma$.
Assuming that a fraction $\epsilon_e$
of total energy goes into the electron acceleration
and a fraction $\epsilon_B$ goes into the magnetic amplification,
we have the typical Lorentz factor of electrons
as $\gamma_m \sim \epsilon_e m_p/m_e$
and the magnetic luminosity as
$\epsilon_B L_{tot} = 4 \pi r^2 c \Gamma^2 B^2/8\pi$
where $r$ is the shock radius and
$L_{tot}$ is the total luminosity.
Since almost all electron energy is radiated, the photon luminosity is
given by $L_{\gamma}=\epsilon_e L_{tot}$.
Combining these relations, we have the peak energy as a function of
$r$, $\Gamma$ and $L_{\gamma}$ as
\begin{equation}
E_{\rm peak} \sim 3~  \epsilon_B^{1/2} \epsilon_e^{3/2}
L_{\gamma,52}^{1/2} r_{13}^{-1} \ {\rm MeV},
\label{eq:Epsyn}
\end{equation}

\noindent where $Q_x=Q/10^x$ in the cgs unit.
In this model, the low energy photon index should not exceed $-2/3$,
and even $-3/2$ if electrons cool fast as expected in GRBs \citep{ghi00}.
 However, many previous observations found flatter low energy
 index than $-2/3$ (e.g., \cite{Preece
1998}; \cite{Ghirlanda 2002}).
 Since our results for GRB 061007 exhibit flatter low energy
 photon index than $-2/3$ for the spectrum of one time interval,
 the photosphere model should also be considered.

In the photosphere model,
the peak energy is identified with the thermal peak
$E_{\rm peak} \sim \Gamma T'$ of the photosphere
under which the outflow energy is internally dissipated and thermalized.
Most opacity of the photosphere
is provided by $e^{\pm}$ pairs or electrons associated with baryons
(e.g., \cite{mi08}).
This model has several advantages
that the peak energy is stabilized even for high radiative efficiency
and
that the low energy photon index can be as hard as the thermal one, $\sim 1$
\citep{th07,ioka07}.
With the Stefan-Boltzmann formula, we have
the comoving energy density of photons as
$a T'^4= L_\gamma/4\pi r^2 c \Gamma^2$, so that
\begin{equation}
E_{\rm peak} \sim  1~ \Gamma_{3}^{1/2} r_{10}^{-1/2} L_{\gamma,52}^{1/4} \ {\rm MeV}.
\label{eq:Epphoto}
\end{equation}

From equations (\ref{eq:Epsyn}) and (\ref{eq:Epphoto}),
the high peak energy above the Yonetoku relation
$E_{\rm peak} \propto L_{\gamma}^{1/2}$ suggests
that the shock radius $r$ is small
and/or the Lorentz factor $\Gamma$ is large.
This means that the fireball is expanding and/or decelerating
in the initial rising of pulses, and we could be directly observing
the fireball dynamics through the evolution of the Yonetoku relation.
Since the shift of the peak energy is about a factor of two,
the radius increases by a factor of $2$-$4$
and/or the Lorentz factor decreases by a factor of $\sim 4$.
Because the time dependence could be different between
the synchrotron shock model $E_{\rm peak} \propto r^{-1} \propto t^{-1}$
and the photosphere model $E_{\rm peak} \propto \Gamma^{1/2} r^{-1/2}
\propto \Gamma^{1/2} t^{-1/2}$,
we could potentially discriminate models
with similar but more detail spectral observations.

\section*{acknowledgments}
We thank Grzegorz Madejski for useful comments.
This work is supported in part
by the Grant-in-Aid from the
Ministry of Education, Culture, Sports, Science and Technology
(MEXT) of Japan, No.18740147, 19047004 (K.I.) and No. 20041001 (M.T.),
and also supported by a special postdoctoral researchers program in RIKEN.

\section*{Appendix}

In this section, we present the detail of the verification processes 
to check whether these outliers in the Figure 8 are real or not from the point of
view of the spectral analysis. At first, we
compare the spectral shape of these outliers with that of main
population as shown in figure \ref{systematic_spec} left. For this comparison, we produce
two integrated spectra. One of them is the spectrum of outliers which are 
summed up among all 8 data points which we defined as the outliers, and
the other is the spectrum of the main population which is the summed up 9 data
points with similar luminosity as the outliers. These selected data
points are shown in the same figure. From this comparison, we find that
the spectral shape of outliers is surely different from that of main
population, and thus we confirm that these outliers do not simply arise from
the uncertainty of the fitting procedures. 
Then, we investigate another possibility that the lack of time 
resolution of the WAM-TRN data causes the behavior that the pulse rising
phase tends to have higher $E^{\rm t}_{\rm peak}$ value. We can only resolve
with 1-sec time scale by the WAM data even if the hard-to-soft 
spectral evolution exists with finer time scale. The $E^{\rm t}_{\rm peak}$
value might be shifted if we average the spectrum out  with 1-sec time
scale which have that hard-to-soft evolution.
In order to confirm this, we performed a Monte Carlo simulation to
reproduce the spectra with hard-to-soft evolution. In this simulation, 
we assume that the $E^{\rm t}_{\rm peak}$ and the luminosity of each spectrum
change, following $E_{\rm peak} \propto L_{\rm iso}^{0.5}$, as in the Yonetoku
relation, to represent the spectral evolution. We then investigate the
differences of the $E^{\rm t}_{\rm peak}$ value between each spectrum and that
of averaged spectrum. As a result, we find that if we average the
spectra with hard-to-soft evolution, the $E^{\rm t}_{\rm peak}$ value becomes
somewhat higher than than of each simulated spectrum, by about a factor
of 1.3. However, for the case of GRB 061007, the difference of the $E^{\rm t}_{\rm
peak}$ value between the outliers and the main population is much larger
than this effect, which is about a factor of 2, and we cannot explain the
behavior of the outliers only by this spectral evolution effect. 

Next, we estimate the systematic uncertainties of these 
time-resolved spectral parameters. 
There are three major systematic uncertainties for 
the WAM data. The first
one is the uncertainty of the detector response matrix. However we
already confirm that this uncertainty should be within 20\% from
the result of time-averaged spectral analysis. Moreover, since we
compare each time-resolved spectrum of only one burst, the uncertainties
of response matrix do not influence our time-resolved analysis. The
second issue is the dependence on the burst intensity. To investigate
this uncertainty, we calculate various types of spectra with same
spectral parameters and different intensities by Monte Carlo simulation.
Figure \ref{systematic} left shows the variation of the spectral parameters as a function
of the burst intensity obtained by these simulated spectra, and we find
that the $E_{\rm peak}$ goes to a lower value as the burst intensity
goes down. However, this variation should be within only 20\% and 
this tendency is inconsistent with that of 
outliers of our time-resolved spectra, i.e. the $E^{\rm t}_{\rm peak}$ value
of outliers becomes larger than major population. 
The third uncertainty is the
variability of the background. Figure \ref{systematic}, right panel, shows a comparison of the
background spectra between pre- and post-burst time intervals, and we 
find that the background spectrum only varies by less than 5\%. We 
find that the $E_{\rm peak}$ value changes only 10\% at maximum even
when we use the background larger by 5\%. This amount of variation cannot
explain the differences of the $E^{\rm t}_{\rm peak}$ between the main
population and the outliers of about a factor of 2.
From these estimation of the systematic effects of our time-resolved
spectral analysis, we conclude that the differences of the main
population and outliers cannot be explained by the systematic
uncertainties and this different population is real for this burst.

\clearpage

\begin{table}[!h]
\small
\caption{Results of the time integrated joint fittings for GRB 061007.}
\label{bestfit_timeavegrb061007}
\begin{center}
\begin{tabularx}{13cm}{clccccc}
\hline
\hline
Model & Det & $\alpha$ & $\beta$ & $E_{\rm peak}$ & C(WAM)$^\dagger$ &$\chi^2/ d.o.f$ \\
\hline
\multicolumn{6}{c}{Interval A}\\
\hline
PL & WAM3 & 2.07$^{+0.06}_{-0.06}$&-&-&-&54/24 \\
& BAT & 1.09$^{+0.05}_{-0.05}$&-&-&-&49/57 \\
& BAT+WAM3 & 1.66$^{+0.04}_{-0.03}$&-&-&1.69$^{+0.21}_{-0.14}$ &570/82 \\
\hline
CPL & WAM3 & 1.13$^{+0.34}_{-0.38}$&-&368$^{+45}_{-37}$&-&15/23 \\
& BAT & 0.98$^{+0.16}_{-0.23}$&-&-&-&48/56 \\
& BAT+WAM3 & 0.92$^{+0.07}_{-0.07}$&-&366$^{+35}_{-31}$&1.07$^{+0.10}_{-0.09}$ &65/81 \\
\hline
Band & WAM3 & -1.02$^{+0.50}_{-0.36}$&-2.99$^{+0.46}_{-7.01}$&358$^{+41}_{-38}$&-&13/22 \\
& BAT & -0.97$^{+0.22}_{-0.16}$&-&-&-&48/55 \\
& BAT+WAM3 & -0.91$^{+0.07}_{-0.07}$&-2.95$^{+0.43}_{-1.72}$&356$^{+37}_{-35}$&1.08$^{+0.10}_{-0.09}$ &62/80 \\
\hline
\hline
\multicolumn{6}{c}{Interval B}\\
\hline
PL & WAM3 & 2.05 &-&-&-&639/24 \\
& BAT & 0.93$^{+0.03}_{-0.03}$&-&-&-&21/57 \\
& BAT+WAM3 & 1.71 &-&-&2.31 &4052/82 \\
\hline
CPL & WAM3 & 0.80$^{+0.12}_{-0.12}$&-&478$^{+16}_{-15}$&-&36/23 \\
& BAT & 0.92$^{+0.03}_{-0.03}$&-&-&-&21/56 \\
& BAT+WAM3 & 0.75$^{+0.03}_{-0.03}$&-&478$^{+15}_{-14}$&1.12$^{+0.05}_{-0.04}$ &65/81 \\
\hline
Band & WAM3 & -0.66$^{+0.18}_{-0.16}$&-3.22$^{+0.25}_{-0.42}$&460$^{+18}_{-18}$&-&22/22 \\
& BAT & -0.92$^{+0.03}_{-0.03}$&-&-&-&21/55 \\
& BAT+WAM3 & -0.74$^{+0.03}_{-0.03}$&-3.31$^{+0.27}_{-0.45}$&467$^{+16}_{-15}$&1.13$^{+0.05}_{-0.04}$ &52/80 \\
\hline
\hline
\multicolumn{6}{c}{Interval A+B}\\
\hline
PL & WAM3 & 2.04 &-&-&-&456/24 \\
& BAT & 0.99$^{+0.03}_{-0.03}$&-&-&-&27/57 \\
& BAT+WAM3 & 1.66 &-&-&1.87&3169/82 \\
\hline
CPL & WAM3 & 0.83$^{+0.13}_{-0.14}$&-&467$^{+18}_{-16}$&-&29/23 \\
& BAT & 0.91$^{+0.03}_{-0.07}$&-&-&-&27/56 \\
& BAT+WAM3 & 0.82$^{+0.03}_{-0.03}$&-&468$^{+17}_{-16}$&1.12$^{+0.05}_{-0.05}$ &61/81 \\
\hline
Band & WAM3 & -0.68$^{+0.20}_{-0.18}$&-3.17$^{+0.26}_{-0.46}$&447$^{+20}_{-19}$&-&18/22 \\
& BAT & -0.98$^{+0.03}_{-0.03}$&-&-&-&26/55 \\
& BAT+WAM3 & -0.81$^{+0.03}_{-0.03}$&-3.27$^{+0.29}_{-0.54}$&456$^{+18}_{-17}$&1.13$^{+0.05}_{-0.05}$ &51/80 \\
\hline
\multicolumn{7}{l}{\scriptsize $\dagger$: constant factor of the WAM against
 the BAT in joint spectral fittings.}
\end{tabularx}
\end{center}
\normalsize
\end{table}


\begin{figure}[!ht]
\begin{center}
\begin{minipage}{7cm}
\hspace{-0.8cm}
\rotatebox{-90}{\resizebox{5.5cm}{!}{\includegraphics{figure/figure3a.ps}}}
\end{minipage}
\begin{minipage}{7cm}
\rotatebox{-90}{\resizebox{5.5cm}{!}{\includegraphics{figure/figure3b.ps}}}
\end{minipage}
\caption{Left panel shows the relation between T$_{90}$ duration and hardness ratio (100-300 keV to 50-100 keV). The dots show BATSE data
 (\cite{Paciesas 1999}). The value of GRB 061007 is plotted by the red
 circle. Right panel shows the $E_{\rm peak,src}$-$E_{\rm iso}$ relation of
 the previous observations (\cite{Amati 2006}). The solid line shows the 
 power-law best fit of this correlation; $E_{\rm peak,src}$ = 
  95 $\times$ $E_{\rm iso}^{0.49}$.  The dashed line shows the 2 $\sigma$ confidence region. The result of GRB 061007 is indicated by the red circle.}
\label{grb061007_HRandAmati}
\end{center}
\end{figure}

\begin{figure}[t]
\begin{center}
\rotatebox{-90}{\resizebox{6cm}{!}{\includegraphics{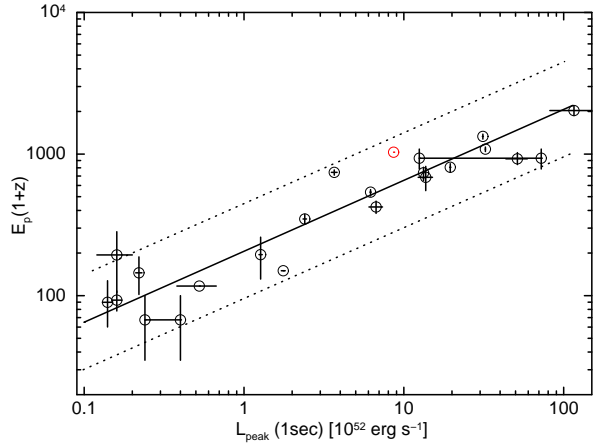}}}
\caption{Relation between $E_{\rm peak}$ in the source frame and peak
 luminosity of 1-sec time scale. Previous result obtained by Yonetoku et
 al. (2004) is shown by black circles, and our result of GRB 061007 is
 shown by red circle. The solid line and dashed-lines show the power-law
 best fit of this relation; $L_{\rm iso} = 2.34 \times 10^{-5} \times
 E_{\rm peak,src}^{2.0}$, and 3 $\sigma$ confidence region.}
\label{grb061007_yonetoku}
\end{center}
\end{figure}

\begin{figure}[t]
\begin{center}
\rotatebox{-90}{\resizebox{6cm}{!}{\includegraphics{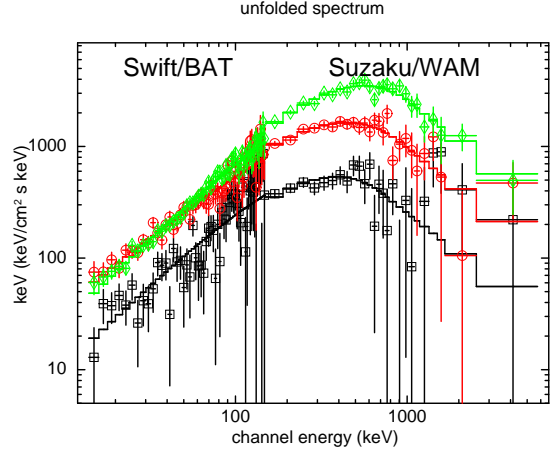}}}
\caption{$\nu F_\nu$ spectra of three time-resolved spectra of GRB
 061007. These spectra are extracted by 1-sec time resolution.
Solid lines represent the best fit
 Band modes. Black squares, red circles, and green diamonds correspond to the
 data of the time region of T=10, 36, 51, respectively. 
}
\label{grb061007_ufu}
\end{center}
\end{figure}

\begin{figure}[!ht]
\begin{center}
\rotatebox{-90}{\resizebox{8cm}{!}{\includegraphics{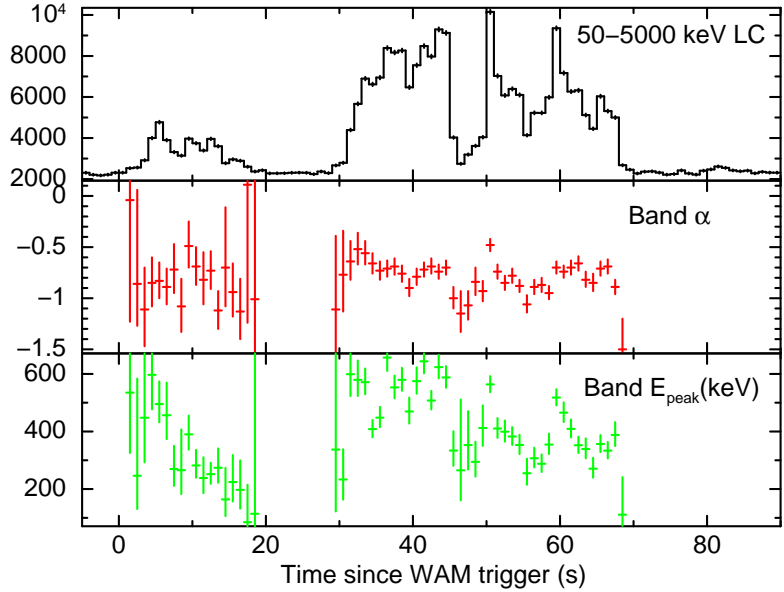}}}
\caption{Time history of the 50-5000 keV count rate (top panel) and the
 Band spectral parameters ($\alpha, E^{\rm t}_{ \rm peak}$), for GRB 061007.}
\label{grb061007_Bandlcpar}
\end{center}
\end{figure}

\begin{table}[!h]	
\begin{center}
\caption{Best-fit parameters of Band model for 58 time resolved spectra of GRB
 061007.}
\label{grb061007_timeresolvedpar}
\small
\begin{tabularx}{14cm}{clccccccc}	
\hline	
\hline	

     &\multicolumn{4}{c}{Band model parameters} & \multicolumn{2}{c}{Intrinsic parameters} \\
\hline
Time &$\alpha$ & $E^{\rm t}_{ \rm peak}$[keV] & C(WAM)$^\dagger$ & $\chi^2/ d.o.f$ & $E^{\rm t}_{ \rm peak,src}$ [keV]& $L^{\rm t}_{ \rm iso}$ [10$^{52}$ erg s$^{-1}$]\\
\hline
2 & -0.04$^{+2.74}_{-1.19}$&534$^{+926}_{-209}$&0.66$^{+1.57}_{-0.48}$ &79/81 &1209$^{+2088}_{-474}$ & 0.62$^{+1.50}_{-0.62}$\\
3 & -0.86$^{+0.92}_{-0.41}$&246$^{+337}_{-115}$&1.79$^{+1.37}_{-0.94}$ &75/81 &557$^{+768}_{-261}$ & 0.23$^{+0.23}_{-0.11}$\\
4 & -1.11$^{+0.41}_{-0.36}$&447$^{+373}_{-155}$&1.42$^{+0.88}_{-0.51}$ &78/81 &1007$^{+847}_{-345}$ & 0.68$^{+0.39}_{-0.24}$\\
5 & -0.85$^{+0.25}_{-0.23}$&596$^{+206}_{-119}$&1.37$^{+0.44}_{-0.32}$ &68/81 &1349$^{+466}_{-270}$ & 1.96$^{+0.58}_{-0.45}$\\
6 & -0.83$^{+0.18}_{-0.17}$&495$^{+78}_{-63}$&1.52$^{+0.36}_{-0.28}$ &88/81 &1119$^{+178}_{-143}$ & 2.56$^{+0.53}_{-0.45}$\\
7 & -0.89$^{+0.18}_{-0.17}$&456$^{+113}_{-82}$&0.99$^{+0.24}_{-0.19}$ &71/81 &1031$^{+256}_{-187}$ & 2.34$^{+0.58}_{-0.46}$\\
8 & -0.72$^{+0.25}_{-0.23}$&269$^{+80}_{-58}$&0.97$^{+0.29}_{-0.22}$ &84/81 &609$^{+182}_{-132}$ & 1.25$^{+0.36}_{-0.27}$\\
9 & -1.08$^{+0.27}_{-0.25}$&265$^{+142}_{-82}$&1.24$^{+0.48}_{-0.33}$ &80/81 &599$^{+321}_{-187}$ & 0.85$^{+0.31}_{-0.21}$\\
10 & -0.49$^{+0.24}_{-0.21}$&390$^{+65}_{-53}$&0.86$^{+0.22}_{-0.17}$ &89/81 &882$^{+148}_{-120}$ & 2.52$^{+0.61}_{-0.49}$\\
11 & -0.69$^{+0.19}_{-0.18}$&282$^{+53}_{-42}$&0.98$^{+0.22}_{-0.18}$ &97/81 &638$^{+122}_{-96}$ & 1.76$^{+0.36}_{-0.30}$\\
12 & -0.82$^{+0.27}_{-0.24}$&238$^{+72}_{-52}$&1.26$^{+0.40}_{-0.29}$ &75/81 &538$^{+163}_{-118}$ & 1.05$^{+0.29}_{-0.22}$\\
13 & -0.73$^{+0.19}_{-0.18}$&251$^{+39}_{-33}$&1.31$^{+0.29}_{-0.23}$ &72/81 &569$^{+89}_{-76}$ & 1.68$^{+0.31}_{-0.26}$\\
14 & -1.12$^{+0.19}_{-0.18}$&274$^{+67}_{-52}$&1.46$^{+0.39}_{-0.29}$ &89/81 &619$^{+152}_{-117}$ & 1.27$^{+0.28}_{-0.22}$\\
15 & -0.70$^{+0.59}_{-0.38}$&163$^{+109}_{-57}$&1.02$^{+0.71}_{-0.37}$ &67/81 &370$^{+249}_{-130}$ & 0.56$^{+0.26}_{-0.16}$\\
16 & -0.94$^{+0.28}_{-0.24}$&224$^{+93}_{-67}$&1.14$^{+0.45}_{-0.30}$ &63/81 &507$^{+212}_{-152}$ & 0.80$^{+0.27}_{-0.20}$\\
17 & -1.13$^{+0.32}_{-0.27}$&197$^{+102}_{-65}$&1.26$^{+0.59}_{-0.38}$ &85/81 &445$^{+232}_{-147}$ & 0.61$^{+0.22}_{-0.15}$\\
18 & 0.11$^{+3.79}_{-1.35}$&85$^{+130}_{-26}$&2.51$^{+2.52}_{-1.57}$ &82/81 &191$^{+295}_{-59}$ & 0.22$^{+0.19}_{-0.06}$\\
19 & -1.01$^{+2.23}_{-0.72}$&114$^{+1462}_{-62}$&0.90$^{+2.55}_{-0.69}$
 &97/81 &257$^{+3333}_{-140}$ & 0.19$^{+0.30}_{-0.08}$\\
\hline
\multicolumn{8}{l}{\scriptsize $\dagger$: constant factor of the WAM against
 the BAT in joint spectral fittings.}
\end{tabularx}
\end{center}
\end{table}

\begin{table}[!h]	
\begin{center}
\caption{Continued.}
\label{grb061007_timeresolvedpar2}
\small
\begin{tabularx}{14cm}{clccccccc}	
\hline	
\hline	

     &\multicolumn{4}{c}{Band model parameters} & \multicolumn{2}{c}{Intrinsic parameters} \\
\hline
Time &$\alpha$ & $E^{\rm t}_{ \rm peak}$[keV] & C(WAM)$^\dagger$ & $\chi^2/ d.o.f$ & $E^{\rm t}_{ \rm peak,src}$ [keV]& $L^{\rm t}_{ \rm iso}$ [10$^{52}$ erg s$^{-1}$]\\
\hline
30 & -1.11$^{+0.72}_{-0.47}$&337$^{+1847}_{-215}$&1.37$^{+1.43}_{-0.63}$ &65/81 &764$^{+4148}_{-489}$ & 0.36$^{+0.50}_{-0.18}$\\
31 & -0.77$^{+0.43}_{-0.36}$&233$^{+105}_{-70}$&1.49$^{+0.80}_{-0.50}$ &92/81 &527$^{+237}_{-159}$ & 0.46$^{+0.21}_{-0.14}$\\
32 & -0.64$^{+0.20}_{-0.18}$&599$^{+97}_{-75}$&1.24$^{+0.29}_{-0.23}$ &92/81 &1354$^{+220}_{-171}$ & 2.97$^{+0.67}_{-0.55}$\\
33 & -0.52$^{+0.16}_{-0.15}$&579$^{+67}_{-56}$&1.07$^{+0.19}_{-0.16}$ &75/81 &1310$^{+152}_{-127}$ & 5.23$^{+0.87}_{-0.75}$\\
34 & -0.56$^{+0.12}_{-0.11}$&571$^{+47}_{-42}$&1.21$^{+0.17}_{-0.14}$ &81/81 &1292$^{+108}_{-95}$ & 6.49$^{+0.85}_{-0.76}$\\
35 & -0.66$^{+0.10}_{-0.10}$&408$^{+31}_{-28}$&1.27$^{+0.15}_{-0.13}$ &104/81 &923$^{+71}_{-64}$ & 5.00$^{+0.54}_{-0.49}$\\
36 & -0.73$^{+0.09}_{-0.09}$&448$^{+37}_{-33}$&1.21$^{+0.14}_{-0.12}$ &71/81 &1013$^{+84}_{-75}$ & 5.85$^{+0.63}_{-0.57}$\\
37 & -0.71$^{+0.08}_{-0.08}$&656$^{+52}_{-46}$&1.06$^{+0.10}_{-0.09}$ &88/81 &1485$^{+119}_{-105}$ & 10.45$^{+1.03}_{-0.94}$\\
38 & -0.69$^{+0.08}_{-0.08}$&553$^{+41}_{-37}$&1.18$^{+0.12}_{-0.11}$ &74/81 &1250$^{+93}_{-84}$ & 8.39$^{+0.83}_{-0.75}$\\
39 & -0.76$^{+0.08}_{-0.08}$&579$^{+41}_{-37}$&1.26$^{+0.12}_{-0.11}$ &80/81 &1310$^{+94}_{-85}$ & 8.36$^{+0.79}_{-0.73}$\\
40 & -0.90$^{+0.09}_{-0.08}$&469$^{+48}_{-42}$&1.37$^{+0.15}_{-0.14}$ &69/81 &1061$^{+109}_{-95}$ & 4.75$^{+0.52}_{-0.47}$\\
41 & -0.79$^{+0.08}_{-0.08}$&574$^{+48}_{-43}$&1.23$^{+0.12}_{-0.11}$ &84/81 &1299$^{+109}_{-98}$ & 7.25$^{+0.73}_{-0.66}$\\
42 & -0.72$^{+0.07}_{-0.07}$&643$^{+46}_{-42}$&1.13$^{+0.10}_{-0.09}$ &103/81 &1456$^{+104}_{-95}$ & 9.99$^{+0.91}_{-0.84}$\\
43 & -0.69$^{+0.08}_{-0.07}$&507$^{+33}_{-30}$&1.28$^{+0.12}_{-0.11}$ &103/81 &1147$^{+75}_{-70}$ & 7.34$^{+0.67}_{-0.61}$\\
44 & -0.74$^{+0.07}_{-0.07}$&623$^{+42}_{-39}$&1.29$^{+0.11}_{-0.10}$ &81/81 &1410$^{+96}_{-88}$ & 10.10$^{+0.86}_{-0.79}$\\
45 & -0.70$^{+0.07}_{-0.07}$&587$^{+39}_{-35}$&1.25$^{+0.11}_{-0.10}$ &66/81 &1329$^{+88}_{-81}$ & 9.70$^{+0.82}_{-0.76}$\\
46 & -1.00$^{+0.11}_{-0.10}$&333$^{+67}_{-53}$&1.25$^{+0.19}_{-0.17}$ &78/81 &753$^{+152}_{-119}$ & 1.90$^{+0.31}_{-0.25}$\\
47 & -1.15$^{+0.22}_{-0.18}$&265$^{+246}_{-104}$&1.01$^{+0.40}_{-0.28}$ &70/81 &596$^{+562}_{-233}$ & 0.63$^{+0.23}_{-0.17}$\\
48 & -1.07$^{+0.14}_{-0.14}$&352$^{+117}_{-84}$&1.26$^{+0.29}_{-0.23}$ &72/81 &797$^{+267}_{-191}$ & 1.07$^{+0.26}_{-0.20}$\\
49 & -0.84$^{+0.14}_{-0.13}$&294$^{+70}_{-51}$&1.28$^{+0.24}_{-0.20}$ &86/81 &662$^{+161}_{-112}$ & 1.34$^{+0.27}_{-0.20}$\\
50 & -0.93$^{+0.10}_{-0.10}$&412$^{+78}_{-62}$&1.24$^{+0.19}_{-0.16}$ &81/81 &931$^{+178}_{-140}$ & 2.06$^{+0.27}_{-0.29}$\\
51 & -0.48$^{+0.06}_{-0.06}$&563$^{+28}_{-27}$&1.29$^{+0.10}_{-0.09}$ &62/81 &1274$^{+65}_{-61}$ & 10.94$^{+0.86}_{-0.79}$\\
52 & -0.74$^{+0.07}_{-0.07}$&410$^{+36}_{-32}$&1.09$^{+0.09}_{-0.08}$ &95/81 &927$^{+83}_{-73}$ & 6.11$^{+0.55}_{-0.49}$\\
53 & -0.85$^{+0.07}_{-0.07}$&399$^{+38}_{-34}$&1.29$^{+0.12}_{-0.11}$ &52/81 &903$^{+87}_{-78}$ & 4.33$^{+0.41}_{-0.37}$\\
54 & -0.78$^{+0.07}_{-0.07}$&382$^{+32}_{-29}$&1.27$^{+0.11}_{-0.10}$ &60/81 &865$^{+73}_{-66}$ & 4.65$^{+0.41}_{-0.37}$\\
55 & -0.88$^{+0.06}_{-0.06}$&353$^{+36}_{-31}$&1.13$^{+0.10}_{-0.09}$ &83/81 &798$^{+81}_{-71}$ & 4.59$^{+0.41}_{-0.37}$\\
56 & -1.06$^{+0.09}_{-0.08}$&255$^{+50}_{-40}$&1.16$^{+0.16}_{-0.14}$ &80/81 &576$^{+113}_{-90}$ & 2.13$^{+0.28}_{-0.24}$\\
57 & -0.89$^{+0.07}_{-0.07}$&306$^{+36}_{-31}$&1.14$^{+0.11}_{-0.10}$ &63/81 &693$^{+83}_{-71}$ & 3.44$^{+0.34}_{-0.30}$\\
58 & -0.87$^{+0.07}_{-0.07}$&287$^{+32}_{-28}$&1.16$^{+0.12}_{-0.11}$ &72/81 &650$^{+74}_{-63}$ & 3.23$^{+0.31}_{-0.27}$\\
59 & -0.95$^{+0.07}_{-0.06}$&354$^{+37}_{-32}$&1.27$^{+0.11}_{-0.10}$ &69/81 &801$^{+85}_{-74}$ & 4.10$^{+0.37}_{-0.33}$\\
60 & -0.70$^{+0.06}_{-0.06}$&517$^{+29}_{-27}$&1.40$^{+0.10}_{-0.09}$ &81/81 &1171$^{+65}_{-61}$ & 8.69$^{+0.62}_{-0.58}$\\
61 & -0.74$^{+0.06}_{-0.06}$&465$^{+35}_{-32}$&1.32$^{+0.11}_{-0.10}$ &77/81 &1052$^{+80}_{-73}$ & 5.72$^{+0.49}_{-0.45}$\\
62 & -0.70$^{+0.07}_{-0.07}$&409$^{+33}_{-29}$&1.41$^{+0.13}_{-0.12}$ &106/81 &925$^{+74}_{-68}$ & 4.16$^{+0.38}_{-0.35}$\\
63 & -0.66$^{+0.07}_{-0.07}$&352$^{+29}_{-26}$&1.32$^{+0.12}_{-0.11}$ &56/81 &796$^{+66}_{-60}$ & 4.24$^{+0.38}_{-0.34}$\\
64 & -0.82$^{+0.08}_{-0.07}$&339$^{+37}_{-32}$&1.27$^{+0.13}_{-0.12}$ &63/81 &767$^{+83}_{-73}$ & 3.07$^{+0.32}_{-0.28}$\\
65 & -0.85$^{+0.09}_{-0.08}$&270$^{+37}_{-31}$&1.19$^{+0.14}_{-0.13}$ &72/81 &613$^{+85}_{-72}$ & 2.34$^{+0.27}_{-0.23}$\\
66 & -0.71$^{+0.07}_{-0.07}$&356$^{+31}_{-28}$&1.20$^{+0.11}_{-0.10}$ &73/81 &806$^{+70}_{-63}$ & 4.31$^{+0.39}_{-0.35}$\\
67 & -0.69$^{+0.07}_{-0.07}$&333$^{+30}_{-27}$&1.27$^{+0.12}_{-0.11}$ &91/81 &754$^{+68}_{-61}$ & 3.30$^{+0.32}_{-0.28}$\\
68 & -0.89$^{+0.07}_{-0.07}$&388$^{+44}_{-38}$&1.21$^{+0.12}_{-0.11}$ &91/81 &877$^{+100}_{-87}$ & 3.25$^{+0.35}_{-0.31}$\\
69 & -1.50$^{+0.30}_{-0.20}$&110$^{+130}_{-38}$&1.56$^{+0.97}_{-0.56}$ &65/81 &251$^{+294}_{-89}$ & 0.38$^{+0.14}_{-0.08}$\\
\hline
\multicolumn{8}{l}{\scriptsize $\dagger$: constant factor of the WAM against
 the BAT in joint spectral fittings.}
\end{tabularx}
\normalsize
\end{center}
\end{table}

\begin{figure}[!t]
\begin{center}
\rotatebox{-90}{\resizebox{6cm}{!}{\includegraphics{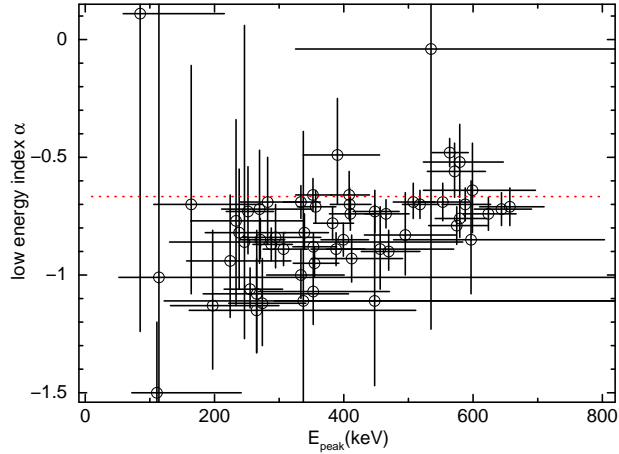}}}
\caption{The relation between observed peak energy $E^{\rm t}_{ \rm peak}$ and
 low energy photon index $\alpha$. The limit of $\alpha$ value predicted
 by the synchrotron emission model ($\alpha=-2/3$) is indicated by
 horizontal dashed-line. Upper region above this line is not allowed by
 the standard synchrotron emission model.}
\label{Epalpha}
\end{center}
\end{figure}

\begin{figure}[!t]
\begin{center}
\rotatebox{-90}{\resizebox{6cm}{!}{\includegraphics{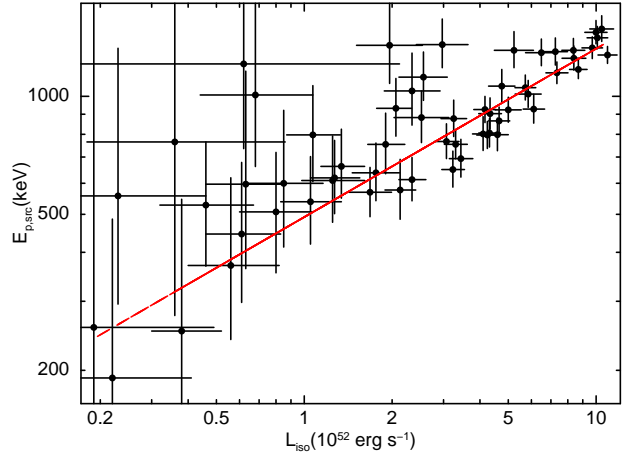}}}
\caption{The relation between isotropic equivalent luminosity $L_{\rm
 iso}$ and the source frame $E^{\rm t}_{ \rm peak}$ measured from each 1-sec
 time-resolved spectra as shown in table \ref{grb061007_timeresolvedpar} and 
\ref{grb061007_timeresolvedpar2}. The power law bet fit
 model is also shown by red dotted line.}
\label{EpLiso}
\end{center}
\end{figure}

\begin{figure}[!t]
\begin{center}
\rotatebox{-0}{\resizebox{6cm}{!}{\includegraphics{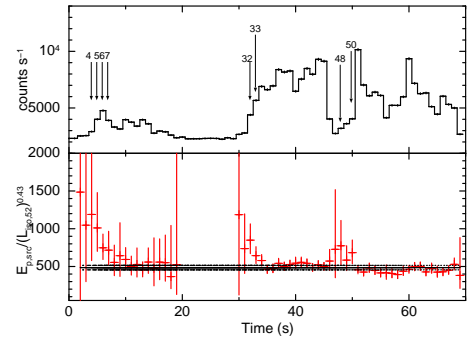}}}
\caption{The light curve (upper panel) and $E^{\rm t}_{ \rm peak}$/$L_{\rm
 iso}^{0.43}$ as the function of time (lower panel). Vertical arrows in
 top panel represent the outliers from the best-fit power law model 
in $E^{\rm t}_{ \rm peak}$-$L^{\rm t}_{ \rm iso}$ plot.  These outliers are located above the 3
 $\sigma$ limit of the best-fit constant parameter in the lower
panel. As seen, all outliers belong to
 initial rising phase of each pulse.  The best-fit constant parameter 
and 3 $\sigma$ confidence level
 are also shown by horizontal solid and dashed lines in lower panel, respectively.}
\label{EpoverLiso045}
\end{center}
\end{figure}

\begin{figure}[!t]
\begin{center}
\rotatebox{-90}{\resizebox{6cm}{!}{\includegraphics{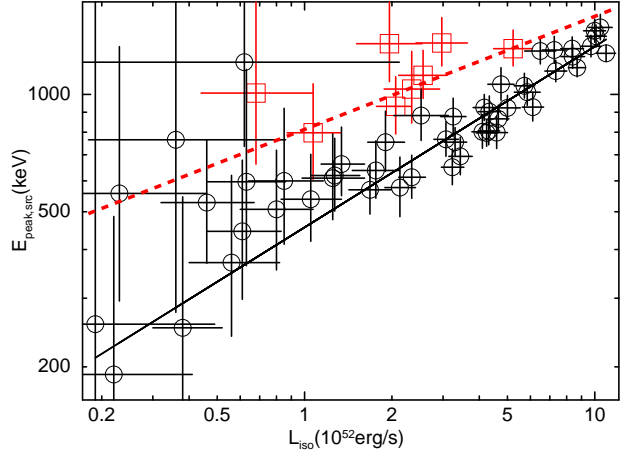}}}
\caption{Same as Figure \ref{EpLiso} but divided into two
 populations; main population (black circles) and outliers (red
 squares). 
The power law best-fit model for each population are also shown by black
 solid
 and red dashed lines.}
\label{EpLiso_fortwopop}
\end{center}
\end{figure}

\begin{figure}[t]
\begin{center}
\hspace{-0.8cm}
\rotatebox{-90}{\resizebox{6cm}{!}{\includegraphics{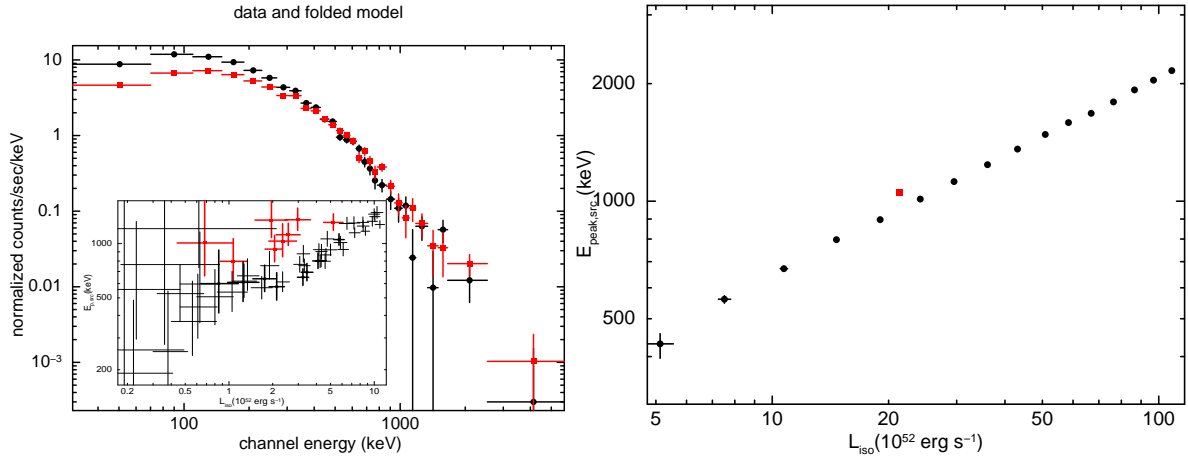}}}
\rotatebox{-90}{\resizebox{6cm}{!}{\includegraphics{figure/figure11b.ps}}}
\caption{Left panel shows a comparison of the integrated spectrum of all
 outliers (red squares) with that of several main population (black
 circles). The selected data are indicated by the thick points in the 
 $E_t{\rm p}-L^{\rm t}_{ \rm iso}$ plane shown in the same figure. The differences
 of the spectral shape for these two population are clearly seen. 
Right panel shows the relation between the $L^{\rm t}_{ \rm iso}$ and the $E_t{\rm
 peak}$ value obtained by the individual simulated spectra (black
 circles), and by the total integrated spectrum (red square). This figure
 indicates that the measured $E^{\rm t}_{ \rm peak}$ value could be higher only by 
factor  about 1.3 even 
if the hard-to-soft evolution following the Yonetoku relation 
 exists at a finer time scale in initial rising phase of each pulse.}
\label{systematic_spec}
\end{center}
\end{figure}

\begin{figure}[t]
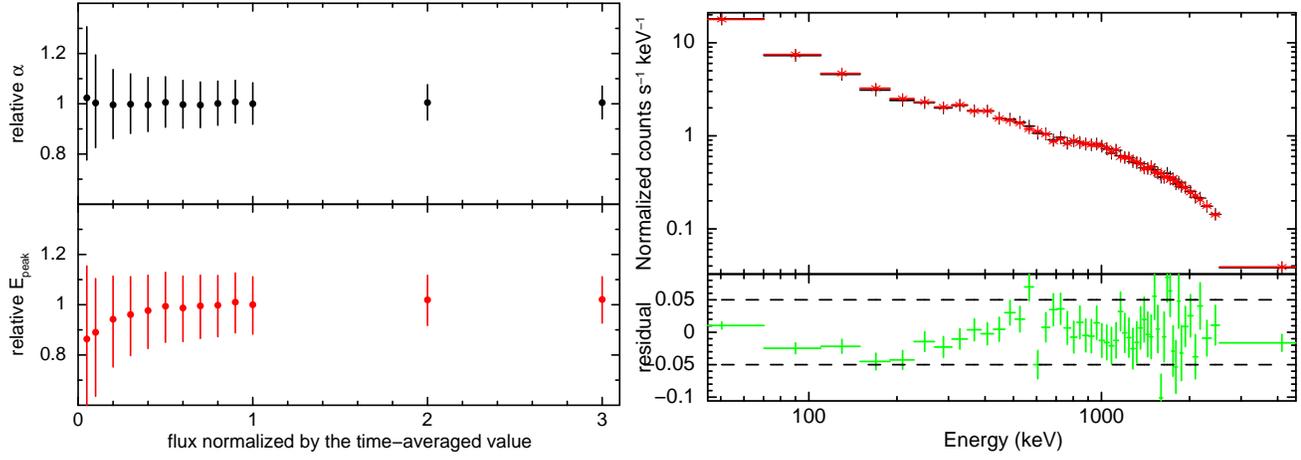

\begin{center}
\hspace{-0.8cm}
\rotatebox{-90}{\resizebox{6cm}{!}{\includegraphics{figure/figure12a.ps}}}
\rotatebox{-90}{\resizebox{6cm}{!}{\includegraphics{figure/figure12b.ps}}}
\caption{Left panel shows a relation between burst intensity and the
 spectral parameters obtained by the Monte Carlo simulation. The burst
 intensity and obtained parameters are normalized by the value obtained
 by the time-averaged spectral analysis. The 20\% unceratinty region is
 indicated by the dashed-lines. This figure indicates that the 
 uncertainty of the spectral parameter depending on the burst intensity 
 should be within 20\%. 
 Right panels shows the comparison of the
 background spectra between before (T$_0$(WAM)-52 s to T$_0$(WAM)-2 s) and
 after (T$_0$(WAM)+87 s to T$_0$(WAM)+137 s) time interval. The residuals
 of these spectra are shown in the lower window. The change
 of the background level is less than 5\% during this observation.} 
\label{systematic}
\end{center}
\end{figure}

\end{document}